\documentclass{llncs}

\usepackage{booktabs}
\usepackage[utf8]{inputenc}%
\usepackage{textcomp}
\usepackage{todonotes}
\usepackage[caption=false,font=footnotesize]{subfig}
\usepackage[capitalize,noabbrev]{cleveref}
\usepackage[hyphens]{url}
\usepackage[printonlyused]{acronym}
\acrodef{PET}{privacy enhancing technology}
\acrodefplural{PET}[PETs]{privacy enhancing technologies}
\acrodef{PIA}{privacy impact assessment}
\acrodefplural{PIA}[PIAs]{privacy impact assessments}
\acrodef{GDPR}{General Data Protection Regulation}
\definecolor{darkgreen}{rgb}{0.0, 0.44, 0.0}
\begin{document}

\title{Privacy Risk Assessment: From Art to Science, By Metrics}

\author{Isabel Wagner \and Eerke Boiten}

\institute{Cyber Technology Institute,\\
De Montfort University, Leicester, UK,\\
\email{\{isabel.wagner, eerke.boiten\}@dmu.ac.uk}}

\maketitle
\begin{abstract}
Privacy risk assessments aim to analyze and quantify the privacy risks associated with new systems.
As such, they are critically important in ensuring that adequate privacy protections are built in.
However, current methods to quantify privacy risk rely heavily on experienced analysts picking the ``correct'' risk level on e.g.\ a five-point scale.
In this paper, we argue that a more scientific quantification of privacy risk increases accuracy and reliability and can thus make it easier to build privacy-friendly systems.
We discuss how the impact and likelihood of privacy violations can be decomposed and quantified, and stress the importance of meaningful metrics and units of measurement.
We suggest a method of quantifying and representing privacy risk that considers a collection of factors as well as a variety of contexts and attacker models. We conclude by identifying some of the major research questions to take this approach further in a variety of application scenarios.
\keywords{privacy risk metrics, privacy impact assessment}
\end{abstract}

\begin{tikzpicture}[overlay, remember picture]
\path (current page.north east) ++(-1.5 ,-1) node[below left, color=darkgreen, text width=19cm] (A){\footnotesize © Springer Nature Switzerland AG 2018. This is the authors' version of the work. It is posted here for your personal use. Not for redistribution.
The final publication is available at link.springer.com, \url{https://doi.org/10.1007/978-3-030-00305-0_17}.};
\node (C)[below=0.2cm of A, color=darkgreen, text width=19cm]{\footnotesize Please cite this work as: Wagner I., Boiten E. (2018) Privacy Risk Assessment: From Art to Science, by Metrics. In: Garcia-Alfaro J., Herrera-Joancomartí J., Livraga G., Rios R. (eds) Data Privacy Management, Cryptocurrencies and Blockchain Technology. DPM 2018, CBT 2018. Lecture Notes in Computer Science, vol 11025. Springer, Cham};
\end{tikzpicture}

\section{Introduction}
A privacy impact assessment (PIA) is the process of identifying and mitigating privacy risks in an existing or planned system.
During a privacy impact assessment, organizations identify possible privacy risks, then quantify and rank these risks, and finally take decisions on whether and how to reduce, remove, transfer, or accept the risks.
``PIA'' also refers to the document produced in this process, and it is generally seen as a {\em living} document in systems development. This is because privacy risks can change over time: as a result of choices made during design and implementation; as a result of evolution of the system and its data governance;
and as a result of developments in processing technology
and availability of related information in the system's environment.

PIAs are an essential component of Privacy by Design~\cite{Cavoukian11}, an approach to dealing with privacy in a proactive rather than reactive way. They have been recommended by national data protection authorities for more than 5 years
already \cite{ICOPIA,CNILPIAold}. In the new European data protection regulation \acsu{GDPR} (\acl{GDPR}) \cite{GDPR}, PIAs (called ``data protection impact assessments'') are mandated for some cases, including surveillance, data sharing, and new technologies.
This is relevant worldwide because of the \ac{GDPR}'s global reach.
As PIAs include consultation with stakeholders, they are also a useful mechanism for obtaining their buy-in in what might otherwise be seen as ``creepy'' data processing processes.

However, the wider application and impact of PIAs may be limited because privacy risk assessments currently rely heavily on experience, analogy and imagination, that is, risk assessment more closely resembles an \textit{art} than a science.
We argue that a more scientific approach to risk assessment can improve the outcomes of privacy impact assessments by making them more consistent and systematic.
Beyond the use to measure and communicate an individual privacy risk, we envision uses of these privacy risk metrics for at least five more purposes: to quantify the effect of privacy controls, to compare the effects of different controls, to analyze trends in privacy risk over time, to compute a system's aggregate privacy risk from its components, and to rank privacy risks.

\textbf{Contributions.}
In this paper, we investigate how to quantify privacy risk systematically with the aim of moving privacy risk assessment from being an art closer to being a science.
We focus on {\em data} driven privacy (i.e. the impact of data decisions, possibly outside the data sphere) because this is the scope of the \ac{GDPR}, currently the strongest driver of PIAs.
In line with the common decomposition of risk into impact and likelihood, we discuss quantification of impact and likelihood separately and suggest possible metrics for each (Sections \ref{sec:impact} and \ref{sec:likelihood}).
We then discuss how metrics for impact and likelihood can be combined to form privacy risk metrics that can be used directly in privacy impact assessments and privacy requirements engineering (Section \ref{sec:combination}).
We illustrate an initial approach to measuring and representing privacy risk in a case study with two typical known privacy risks (Section \ref{casestudy}).
Finally, we highlight open issues in the area of privacy risk quantification and set out an agenda for further research.

\section{State of the Art}

Before we discuss the benefits and building blocks of a more scientific method for quantifying privacy risk, we briefly describe the state of the art in risk assessment, privacy risk assessment, and privacy measurement.

\textbf{Risk assessment.}
Risk is commonly calculated as some function of likelihood and impact.
Several proposals exist to determine the risk of \textit{security} threats, for example the NIST guidelines~\cite{nist2012guide} or the OWASP Risk Rating Methodology~\cite{openwebapplicationsecurityproject2018owasp}.
These are often cited in the privacy literature because security risks can be quite close to privacy risks.
An important difference between security and privacy risk, however, is that harm to individuals is a primary consideration for privacy risk (even if organizations
may translate that into reputational and regulatory risks), whereas it is of secondary importance for security risk.

Both NIST and OWASP rate impact and likelihood on Likert scales, e.g. from ``very low'' to ``very high'', with no clear guidelines on how to determine the position on this scale.
For example, the NIST guidelines \cite{nist2012guide} list examples of adverse impacts, such as harm to operations, assets, or individuals, and explain how the expected extent of each impact should be mapped to the Likert scale: ``significant'' financial loss, for example, is a moderate impact, while ``major'' financial loss is high impact.
Likelihood is split into the likelihood that a threat event occurs, and the likelihood that an adverse impact results from the threat event.
The ratings for likelihood and impact are then combined according to a table that indicates the resulting risk rating for each combination of the separate Likert scale ratings.
For example, ``low'' impact and ``very high'' likelihood result in a ``low'' overall risk.
These impact and likelihood ratings are subjective, i.e., they may be rated differently by different people, and the resulting risk ratings may not be accurate or reliable.
In addition, while these tables allow to distinguish between the lowest and highest risks, they only give a partial ordering of risks.
For example, it may not be possible to decide the ordering of one risk with low impact and high likelihood and another with high impact and low likelihood.

\textbf{Privacy risk assessment.}
The OWASP top-10 list of privacy risks in web applications \cite{stahl2017owasp} ranks privacy risks by their ratings for impact and likelihood.
Likelihood is measured as the frequency with which the risk occurs in existing websites (determined via a survey of web developers and privacy/security experts), with a score of 0 indicating under 25\%, and 3 indicating 75\% or above.
Impact is measured in five dimensions as limited (1), considerable (2), or devastating (3): two dimensions for organizational impact (reputation, finance) and three dimensions for impact on individuals (reputation, finance, freedoms). The final impact score is the average of the five scores.

Recently superseded guidance on privacy risk management by the French data protection regulator CNIL \cite{CNILPIAnew} assesses risk {\em severity}, which is based on the possible {\em prejudicial effects} -- similar to impact -- and on the level of {\em identifiability} of data.
The latter includes aspects of impact, i.e., the loss of highly identifiable data is more impactful, as well as aspects of likelihood, i.e., the ease of exploiting a data loss as a privacy attack depends on the level of identifiability of the targets in the data set.
Albakri et al.\ \cite{Albakri2018} employ this notion to abstract from attackers' motivation and capacity, by assessing both privacy and security risks on the basis of exploitability rather than likelihood.

Several bodies have published lists of known privacy risks, for example data protection authorities \cite{CNILPIAbases}, researchers \cite{deng2011privacy}, and regulators.
Although these lists can serve as starting points for privacy impact assessments, they typically do not include a quantification or ranking of specific privacy risks.

\textbf{Privacy measurement.}
Most privacy metrics that have been proposed in the literature \cite{wagner2018technical} focus on measuring the amount of privacy that a \acl{PET} can provide against some adversary, for example expressed as the adversary's error, uncertainty, or information gain.
Some privacy metrics focus on the adversary's success rate and may thus be suitable to quantify the likelihood of a privacy violation (see \cref{sec:likelihood}).
Very few privacy metrics measure risk directly, for example, the privacy score in social networks \cite{liu2010framework} is computed as the sensitivity of profile items multiplied by their visibility.
However, this metric has limited applicability because of its focus on social networks, and because it  does not consider harm to individuals.

\section{Benefits and Building Blocks for Privacy Risk Metrics}

We see four important benefits that can be achieved through the increased accuracy and reliability of a more scientific and systematic way of measuring privacy risk.
First, when building new systems, risk metrics could allow to compare the risks associated with different ways of building the system.
In particular, for systems that are composed of smaller building blocks, risk could be measured on the level of building blocks, and composition rules would allow to compute the overall risk.
In effect, such risk metrics allow to rationalize and substantiate decisions about how systems that affect privacy are built and evaluated.

Second, risk metrics are also needed in privacy requirements engineering \cite{deng2011privacy}, which is a similar process to \ac{PIA}, but with the goal of deriving formal privacy requirements and identifying suitable protections in the form of privacy-enhancing technologies.
The privacy requirements engineering process can identify many risks and thus needs a way to prioritize risks.
For example, the LINDDUN method~\cite{deng2011privacy} uses risk scores, but does not state specifically how these scores should be determined. %

Third, risk metrics can also allow to set thresholds for when the regulator needs to be consulted (e.g., as per \ac{GDPR} guidance by the UK's Information Commissioner's Office \cite{ico2018guide}), and thresholds for when privacy risks are too high to permit data collection or processing.
Thresholds can also play an important role in organizations' decisions to {\em accept} certain risks -- for example, in large
organizations often risks need to associated with million dollars' damages before they warrant the board's attention. Even vaguely defined metrics can support a {\em triage} process on an identified collection of risks to determine the risks' priorities based on their severities.

Fourth, companies that offer cyber insurance benefit from accurate risk metrics to correctly determine insurance premiums. With their past experiences of incidents that they have already paid given amounts out on, there is no doubt that they already hold the largest vault of monetary valuations of privacy risks, but are unlikely to share this, for commercial reasons.

\textbf{Building blocks for a more scientific risk quantification.}
An important foundation of a more scientific approach is the ability to measure and predict the relevant quantities, i.e. the  likelihood (Section~\ref{sec:likelihood}) and impact (Section~\ref{sec:impact}) of privacy violations.

\begin{figure}
 \centering
 \includegraphics[width=\textwidth]{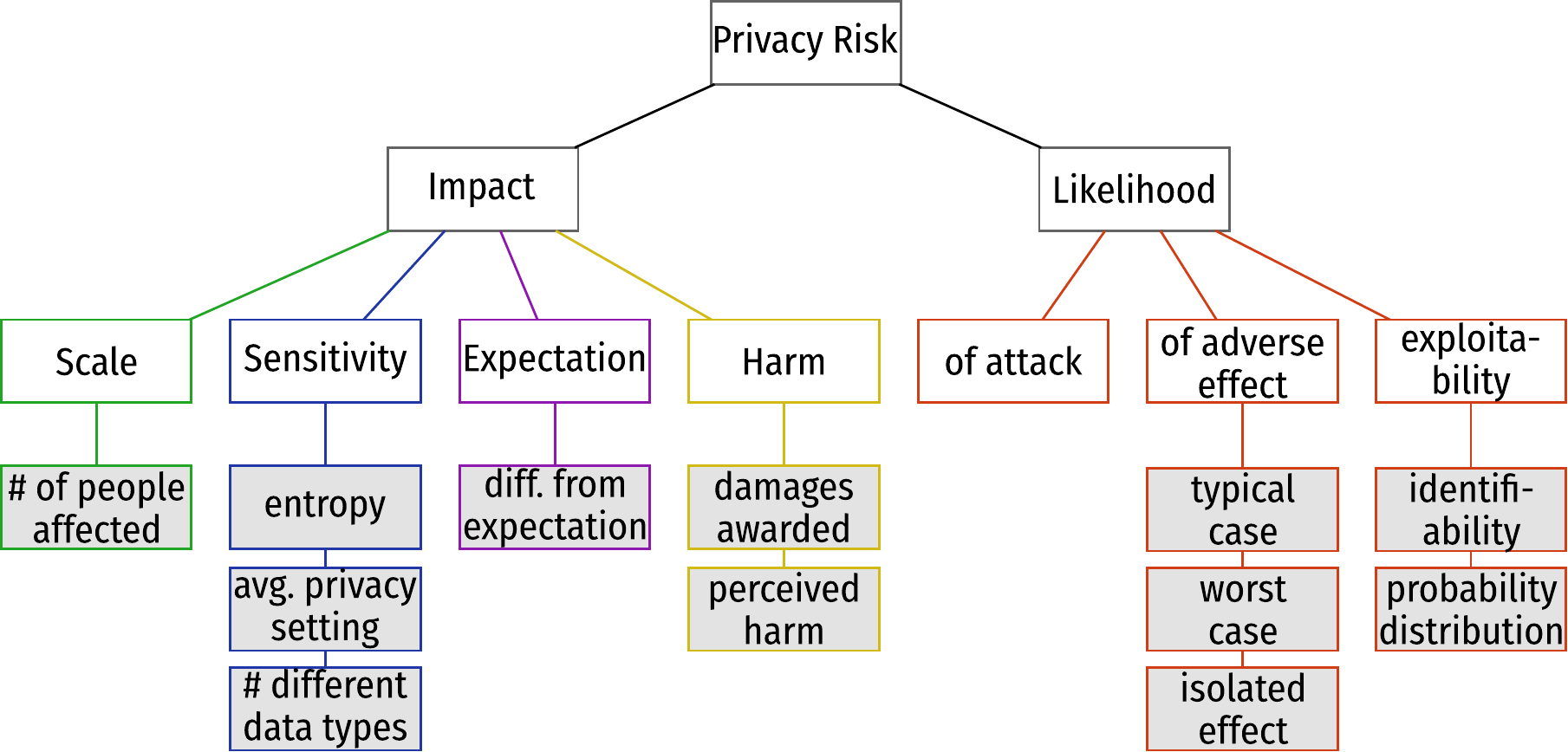}
 \caption{Components of privacy risk.}
 \label{fig:risk-tree}
\end{figure}

To make the measurement of privacy risk more systematic, we decompose the impact and likelihood of privacy risk into more fine-grained components.
As Figure~\ref{fig:risk-tree} shows, we decompose impact into the
four components scale, sensitivity, expectation, and harm, and decompose likelihood into the likelihoods of attack, of adverse effect, and exploitability. 
Because these components are more specific than the high-level concepts of impact and likelihood, it should be easier to find meaningful metrics for them.

However, measuring these quantities is complicated by the fact that the measurement necessarily relies on information known in the present.
Future data sharing or future technologies available to adversaries, such as advanced re-identification algorithms, can significantly increase the privacy risk (but typically do not decrease the risk).
Function creep -- the repurposing of collected data with the intent of realizing new functions -- is also associated with an increase of privacy risk.
Any present-day measurement of privacy risk should therefore be treated as a lower boundary on the real privacy risk.

Units of measurement are important to make risk metrics more understandable and manageable. Risks measured using the same unit can be meaningfully aggregated, for example when computing the total privacy risk from contributing risk factors, and directly compared, for example when considering different technical alternatives or when triaging and prioritizing risks. When units differ, or when there is no unit at all (e.g., in Likert scales), such operations become more difficult or fundamentally dubious.

In business, financial value may be acceptable as the ultimate unit which is used to quantify direct costs -- even reputation and human lives. However, certainly the public sector does not operate
on a competitive or financial basis, and may prefer units that more closely relate to the concept of privacy risk.

\section{Impact Quantification}
\label{sec:impact}

To make the measurement of privacy impact more systematic, privacy impact metrics should be based on four key components (see Figure~\ref{fig:risk-tree}): scale, sensitivity, user expectations, and harm.
Because of the intangible nature of some of these components, we expect that their quantification will have to use proxy measures instead of measuring the component directly.

\subsection{Scale}
The scale of a privacy violation roughly corresponds to the number of people potentially affected by it. Everything else being equal, a violation that affects one person is less severe than one that affects a hundred.
This scale of privacy violations is widely reported in the news when privacy breaches become public.
For example, there have been prominent instances of companies underestimating the scale of privacy breaches when they are first reported, possibly to reduce negative impact on their reputation and hence on their share value.
This underlines the fact that companies treat the scale of a privacy violation as a meaningful metric.

\subsection{Sensitivity}
The sensitivity of the affected data indicates the type and extent of possible harm to individuals.
The sensitivity of data is not necessarily fully aligned with the \ac{GDPR}'s categories of \textit{personal data} and \textit{special category data} -- credit card data are classified as personal data, but can cause direct financial harm, whereas trade union membership is classified as special category data, but its exposure would not be seen as harmful in many countries.

Importantly, if the privacy of more than one type of data is breached, then the overall sensitivity may be higher than a linear combination of individual sensitivities.
For example, the information that a given person was at a location (e.g.\ a celebrity at a nightclub) may not be that sensitive, but the combination of that information with another
person being at the same location at the same time may produce
sensitive evidence of a meeting between the two.

The sensitivity of data is thus difficult to quantify.
Metrics from information theory could be used to measure the amount of information (in bits) revealed by a privacy breach; this will often be indicative of the level of identifiability, but \textit{amount} does not fully coincide with \textit{sensitivity}.
Another approach that is useful when users can choose their individual privacy settings is to compute sensitivity 
statistically from the privacy settings of a large number of users \cite{liu2010framework}.

\subsection{Expectation}
The expectation individuals have of how their data will be treated, and how much a privacy violation deviates from this expectation, indicates as how ``creepy'' a privacy violation will be perceived.
For example, users usually expect that their data will be handled according to their personal privacy settings.
Users may also have expectations where their data is stored, for example, the leak of electronic health records from a third party server located in a foreign country would be unexpected because people may not expect that the storage of health records is outsourced abroad.
Depending on social norms, there may also be a reasonable expectation of privacy in public places~\cite{nissenbaum2004privacy}.
The \ac{GDPR} makes it explicit that the legality of data processing may depend on user expectations\footnote{For example, Recital 47 on the legal basis of ``legitimate interest" requires ``taking into consideration the reasonable expectations of data subjects based on their relationship with the controller."}.
An approach to quantify this deviation from expectation may be to first state the expectation in terms of Solove's taxonomy of privacy \cite{solove2006taxonomy}, i.e.\ to state which aspects of information collection, information processing, information dissemination, or invasion are expected by individuals.
Then, a specific privacy violation can be analyzed with respect to the number of aspects that differ from the stated expectation.

\subsection{Harm}
The harm to affected individuals can be financial, but can also be harm to their reputation, harm caused by discrimination, distress, or anxiety, and harm due to breaches of the individual's rights
and freedoms.
These privacy harms are all covered by (European) data protection legislation\footnote{See \ac{GDPR} Recital 75: 
``The risk to the rights and freedoms of natural persons, of varying likelihood and severity, may result from data processing which could lead to physical, material or non-material damage, in particular: where the processing may give rise to discrimination, identity theft or fraud, financial loss, damage to the reputation, loss of confidentiality of personal data protected by professional secrecy, unauthorized reversal of pseudonymization, or any other significant economic or social disadvantage; where data subjects might be deprived of their rights and freedoms or prevented from exercising control over their personal data" \cite{GDPR}} and it was established before the GDPR came in that individuals can sue for damages even where harms are not material \cite{evans2015vidalhall}.

An important contributing factor in this is what has actually happened to the data: has it been exposed,
modified, processed non-transparently, or used to make a decision affecting individuals? If exposed, to whom and what harms could and would they cause, given existing and potential future information available to the receivers of the data?

Similarly to sensitivity, harm may be cumulative. For example, a single data disclosure may not be very harmful on its own, but a series of disclosures over a period of time may finally allow an adversary to link data and cause serious harm.
This also means that it can be hard to attribute privacy harm to a single privacy breach, which may lead to a dissolution of corporate responsibility, especially when privacy breaches occur along the supply chain.

Harm can also encompass organizational harm, for example reputation damage after the discovery of a privacy breach, or financial damage through loss of customers or regulatory fines.

Finally, individuals may have different perceptions of the harm itself, especially non-financial harm.
As a result, harm is difficult to quantify.
A useful proxy measure may be to estimate the amount of damages a court would be likely to grant.
However, not everything can be measured in money, and expressing harm in monetary terms may not do justice to the extent of the harm caused.
In this case, a Likert scale could be used to estimate the extent of each type of harm affected by a privacy breach.

\section{Likelihood Quantification}
\label{sec:likelihood}
Quantifying the likelihood of a privacy violation is somewhat more tangible than quantifying the impact.
Quantifying likelihood is particularly important because most privacy controls affect the likelihood of a privacy violation instead of its impact.
The NIST guide on privacy engineering~\cite{brooks2017introduction} focuses on the likelihood of ``problematic data actions.''
However, we believe that a thorough quantification of likelihood needs to take into account three aspects of likelihood: the likelihood of an attack, the likelihood of an adverse effect, and exploitability.

\subsection{Likelihood of attack}
The likelihood of an attack focuses on the adversary's motivation to cause a privacy violation.
This is very difficult to quantify because it may depend on specific circumstances.
For example, an adversary may be more motivated to breach medical
data privacy when a celebrity has recently been admitted to a specific hospital.
The arrival of the celebrity may even cause a perfectly innocent staff member at the hospital to turn into an adversary who misuses their access to patient records.

Instead of attempting to estimate this likelihood directly, we believe that it is reasonable to assume that a motivated attacker is present (i.e., assume a likelihood of 1), and to focus on quantifying the other aspects of likelihood.

\subsection{Likelihood of adverse effect}
The likelihood that an adverse effect actually materializes can similarly depend on specific circumstances, and adverse effects may be very rare or not easily attributable to a single privacy violation.
The focus on harm to the individual that is required to assess privacy risk means that it is not sufficient to assess the typical case, but that the worst case also needs to be considered.
Therefore, instead of estimating the exact probability distribution for the occurrence of adverse effects, we believe that it is more beneficial to assess privacy impact for three distinct points on this distribution: the impact on the typical user, the impact on the individual who would be affected worst, and the impact that would be caused if the adversary didn't have any additional information, i.e., the impact caused if this was a single, isolated privacy violation.

\subsection{Exploitability}
Exploitability focuses on the adversary's ability to cause a privacy violation.
Specifically, a systematic quantification should focus on the probability that a specific privacy violation occurs against an adversary with specific aims, capabilities and additional knowledge that corresponds with a realistic attack model.
Considering possible adversaries explicitly is necessary to make the likelihood quantification meaningful and highlights the assumptions made during the privacy risk assessment.

An adversary is any party that is interested in private data, whether within the organization that holds the data, a connected organization such as a service provider, or an external third party \cite{eckhoff2018privacy}. 
Privacy risks can exist even in the absence of attacks, for example through human error and accident. Both can be modeled as attacks by non-malicious insider adversaries.
Privacy risks can occur as a collateral effect even if the adversary is not primarily driven by a privacy-related motivation. For example, an adversary targeting critical national infrastructure may gather information for a spear-phishing attack, and in the process cause privacy harms, even though this is not the primary goal.

There is a wide variety of adversary models considered in the literature (see \cite{wagner2018technical} for an overview).
For adversaries that aim to breach privacy it is especially important to consider inference algorithms that allow the adversary to learn private information from public observations as well as the adversary's prior knowledge because combining data types can increase both likelihood and impact of a privacy breach. 

An important factor in exploitability is identifiability: many privacy attacks are based on knowledge of sensitive information about an identified person. A re-identification attack, in itself an abstract privacy attack, can be the essential stepping stone in this, for example starting from ``anonymized information or ``big data''. Quantification of re-identification risk is difficult \cite{Boiten2016}, not least because there may be large differences between the possibilities of re-identifying a specific individual (such as Governor Weld by Latanya Sweeney \cite{sweeney2002kanonymity}), any individual of choice, or all individuals in a given data set. The 
\ac{GDPR} \cite[Recital 26]{GDPR} nevertheless requires an explicit assessment of what an adversary may ``reasonably likely'' use in attempting to re-identify information.

The result of modeling possible adversaries is a set of probability distributions that indicate how likely it is for each adversary to succeed in breaching privacy.

\section{Privacy Risk Metrics}
\label{sec:combination}
Similarly to security risk metrics, a privacy risk metric could be defined as a combination of metrics for impact and likelihood of privacy violations.
However, our discussion in the previous sections has shown that both the impact and the likelihood of privacy risk are composed of several components that are not easily integrated.
For impact, using our suggested metrics above, we would need to combine the number of people affected, the differences in user expectations, bits of information revealed, and the (monetary equivalent of) harm to individuals.
Ideally, this combination should result in a metric with a meaningful unit, and not just an arbitrary number.
For likelihood, we need to consider both the likelihood of adverse effects and the exploitability for different kinds of adversaries.
As a result, the typical method of adding or multiplying Likert scores does not appear suitable for privacy risk.

In the Introduction, we argued that privacy risk metrics are needed for five purposes: to quantify the effect of privacy controls, to compare the effects of different controls, to analyze trends in privacy risk over time, to compute a system's aggregate privacy risk from its components, and to rank privacy risks.
Each of these purposes has a minimal requirement for the scale of measurement~\cite{stevens1946theory} used by the privacy risk metric.
For example, we need at least an ordinal scale to rank privacy risks, and a ratio scale to analyze aggregate privacy risk in complex systems.
To analyze trends in privacy risk and to compare different privacy controls, an ordinal scale is strictly speaking sufficient, but may not be fine-grained enough to give meaningful or informative results.
We show which scale of measure is required to support each of the five purposes in Table~\ref{tab:scale-vs-purpose}.

\begin{table}
\centering
\caption{Measurement scales required for different purposes of privacy risk metrics}
\label{tab:scale-vs-purpose}
\renewcommand{\arraystretch}{1}
\begin{tabular}{ll}
\toprule
Purpose & Scale of measure \\
\midrule
Effectiveness of privacy controls & Ordinal  \\
Comparison of privacy controls & Ordinal  \\
Trends in privacy risk & Ordinal \\
Calculation of system risk from components \hspace{.5cm} & Ratio \\
Ranking of privacy risks & Ordinal \\
\bottomrule
\end{tabular}
\end{table}

We can see that Likert scores (ordinal, but coarse-grained) 
can be sufficient for some purposes.
However, they are not suitable to analyze the aggregate privacy risk in a complex system, and they are not desirable because, as we have argued, they depend on subjective judgment and may therefore differ depending on who is performing the risk assessment.
In some cases, however, it seems unavoidable to use an ordinal scale, for example to express that an individual's freedoms have been infringed, or the level of distress experienced by an individual.

In these cases, it is unclear how two or more ordinal measures, e.g., for different types of harm, should be combined because the commonly used operations -- addition and multiplication -- are not defined for ordinal scales~\cite{stevens1946theory}.
The usual method of adding or multiplying impact and likelihood scores assigns numerical scores to the levels on the ordinal scale, thus creating a false sense of an interval or ratio scale, for which addition or multiplication would be permitted.
To achieve a clean combination of impact and likelihood metrics, we suggest to measure the individual components separately and combine them visually.
For example, as shown in the case studies in Section \ref{casestudy} (Figs.~\ref{fig:case1} and \ref{fig:case2}), the impact metrics can be combined in a radar plot, and the likelihoods for each adversary type can be indicated with probability density functions or summarized in box plots.
This approach respects the essential multidimensionality of privacy risk and allows to choose appropriate scales for each type of impact.
For example, employment-related harms could be assessed using a 5-point Likert scale ranging from``annoying day'' to ``off with stress'' to ``fired/end of career,'' whereas the scale of the privacy violation could be assessed using the number of individuals affected.

\section{Case Study: Privacy Risks in a Flashlight App}\label{casestudy}

To illustrate how a privacy risk assessment can analyze and visualize the components of privacy risk that we have presented so far, we analyze an example application for a mobile device, focusing on two privacy threats from the OWASP top-10 list of privacy risks in web applications \cite{stahl2017owasp}.

We consider a mobile application that allows users to use their phone as a flashlight. During installation, the app has requested permission to geolocate the user \cite{snoopwall2014flashlight}, and during usage the app displays advertisements \cite{meng2016price}.

\subsection{Collection of Data not Required for Primary Purpose}
The threat that an application collects data that is not required for its primary purpose is rated on the OWASP list as the sixth-highest risk, with high impact and very high frequency.

\begin{figure}[t]
 \centering
 \includegraphics[width=\textwidth]{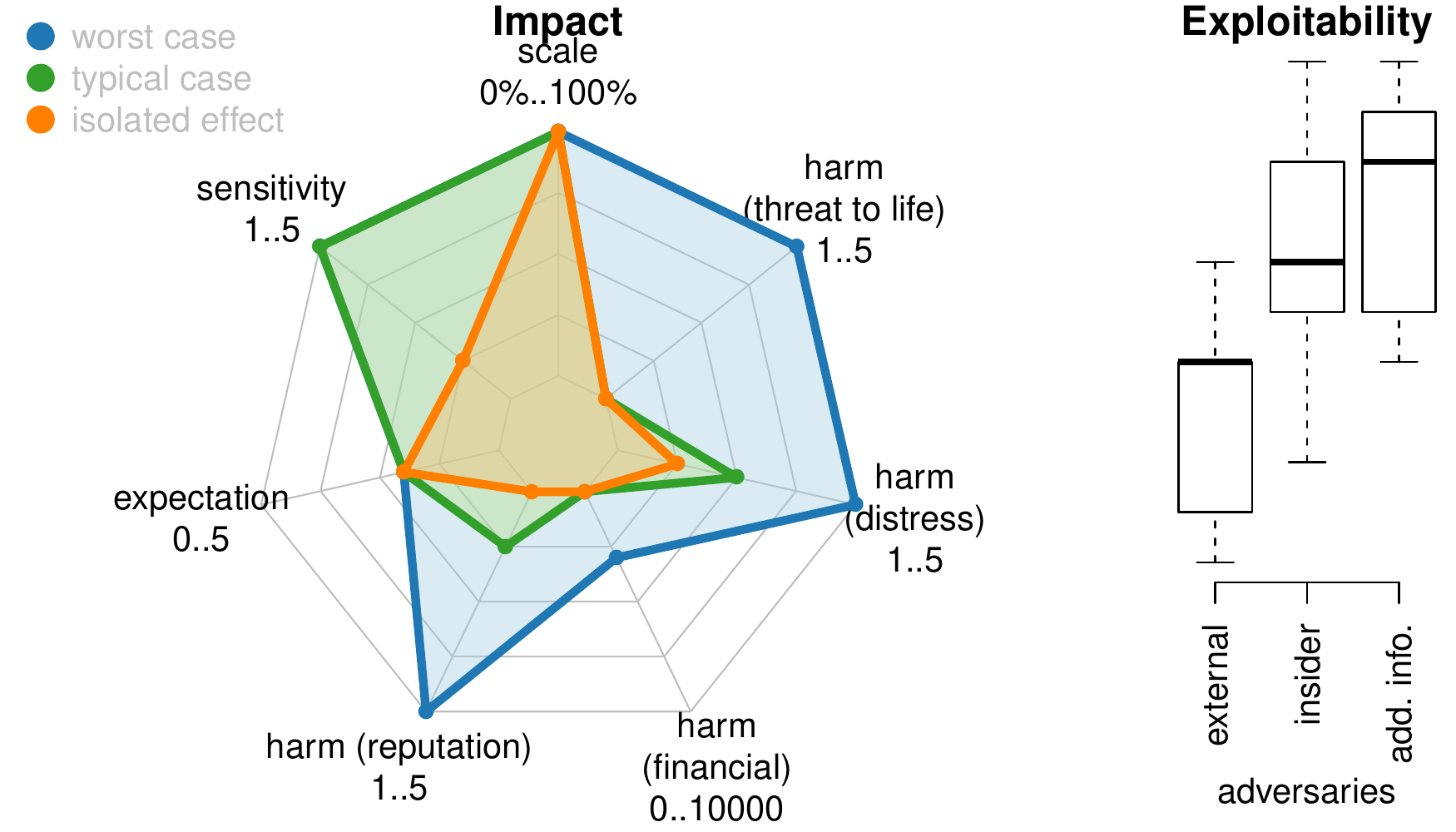}
 \caption{Privacy risk caused by collection of data not required for primary purpose.}
  \label{fig:case1}
\end{figure}

Assuming that the app stores phone identifiers and user locations in a database, a privacy violation can be expected to affect all users of the app.
Correspondingly, the radar plot in Figure~\ref{fig:case1} shows that the scale is 100\% in all cases.

The sensitivity of the data can be classified as very high because geolocation data can allow inferences about behaviors, employment, health, and beliefs. This is especially the case if the app can run in the background and continue to record location data even when not in use.

The expectation of users is that a flashlight app does not collect, process, or share geolocation information~\cite{lin2012expectation}.
However, because the example app does collect and process location data, the expectation differs from reality in two aspects.

The app can cause harm to individual users in terms of reputation damage, financial harm, distress, and a threat to life.

Reputational harm could be caused if, for example, it became public knowledge that an individual regularly located in the red light district.
In the worst case, this could have severe consequences for employment or personal relationships.
The typical user may not have visited particularly sensitive locations, and therefore the typical reputational harm would be much less severe.

Financial harm could be caused if an insurance company obtained the data, determined that some customers were regularly located at a fast-food restaurant, and decided that these customers should be paying higher insurance premiums.
In the worst case, the financial harm could therefore equal the additional yearly cost of insurance to these users, whereas the typical user might not suffer any financial harm.

Harm in terms of distress could be caused if the data revealed a user's home location and patterns of their absence from home.
As a result, users might become afraid of burglaries.
In the worst case, a user might happen to be a stalking victim and may now have to relocate to avoid the stalker.

Harm as a threat to life could be caused if the app was used in a critical environment such as a warzone.
In the worst case, a soldier using the flashlight app -- maybe because the traditional flashlight has failed -- might be targeted by an enemy drone or hand grenade because the app has leaked the soldier's location \cite{perez-pena2018strava}.

We can estimate the likelihood of a privacy violation in terms of exploitability for three cases.
First, if the data has been collected but never used, an external adversary would be limited to sniffing network traffic, which would be a relatively difficult attack.
Second, if the data has been leaked to the public, or if an insider adversary has misused their access privileges, then re-identification attacks are much easier to perform.
Third, the data is most easily exploited if an adversary has additional information that allows to link phone identifiers to real user identities.

\subsection{Sharing of Data with Third Party}
\begin{figure}[t]
 \centering
 \includegraphics[width=\textwidth]{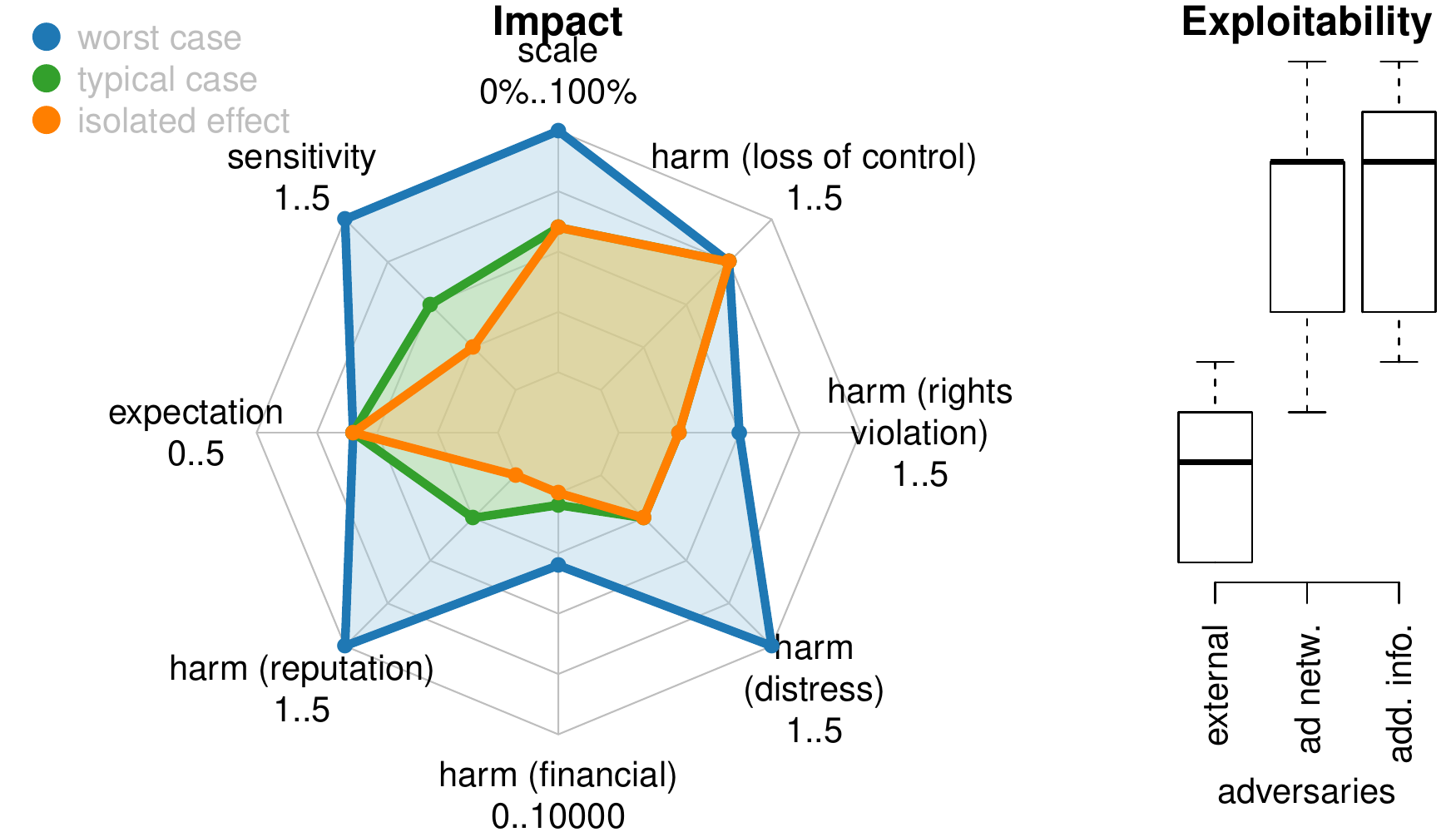}
 \caption{Privacy risk caused by sharing of data with third party.}
  \label{fig:case2}
\end{figure}

The threat that an application shares data with a third party is the seventh-highest risk on the OWASP list, with high impact and high frequency.
We assume that the flashlight app shares data with an advertising network, and that the ad network also uses device fingerprinting to track user activity across all of their applications.
Figure~\ref{fig:case2} visualizes the impact and exploitability for this risk.

Similarly to the first example, the privacy violation can affect all users. However, in some cases, users may have fewer or less interesting interactions with their phones, or may be using ad blockers. In these cases, the scale in terms of the number of affected users would be reduced.

The sensitivity of the data can vary depending on the type of activities that a user performs. In the worst case, these can allow far-reaching inferences about the user's behaviors, purchases, and social life, but we expect that possible inferences in the typical case will be somewhat more limited.

The expectation of users is that a flashlight app does not collect, process, or share device fingerprints, all of which happen in this example. The reality therefore differs from expectation in three aspects.

The primary harms caused by sharing of data with a third party are two abstract types of harm: the violation of basic rights, and the loss of control over data.
The violation of rights is relatively limited, with the exception of children, who are afforded more protection and whose rights are thus affected to a higher degree.
In contrast, the loss of control is fairly severe because the user not only loses control over their data, but is also not informed of the data sharing.

The secondary harms caused by data sharing concern how the third party uses the data, and can be grouped in distress, financial harm, and reputational harm.
Harm in terms of distress can be caused by targeted advertising, which is typically a rather low-level annoyance.
However, device fingerprinting increases an individual's identifiability, which, in the worst case, might lead to the identification of specific individuals as criminals.

Financial harm can be caused by differential pricing, that is, the case when users are offered higher prices for products or services based on their profile.

Harm in terms of reputation damage can be caused, for example, if ads targeted to one user appeared on other users' devices that were falsely attributed to the targeted user, for example the spouse's phone.
In a typical case, this may only ruin birthday surprises, but in the worst case could lead to more severe consequences for relationships or employment.

We can estimate the exploitability of this privacy risk for three types of adversaries.
First, the ad network itself is similar to the insider adversary in the first example, but may be more easily able to exploit the risk because it already has additional data from tracking the user across applications.
Second, an external adversary who can only sniff network traffic would be somewhat more limited than the external adversary in the first example because behavioral profiling data is less easily recognizable than geolocation data.
Finally, an adversary with the ad network's knowledge plus additional information that can be linked to specific users is similarly powerful as in the first example.

\subsection{Discussion}
We have considered two significant and well-known abstract
privacy risks in a concrete scenario, with significant differences on the outcomes in several dimensions of
privacy risk as well as in the adversary profiles.  
Considering the separate factors and adversaries has
led to a deeper understanding and more detailed
representation of the risks. Quantitative information mostly
remained on Likert scales, which means that not all questions we might ask of these scenarios, such as ``which risk is worse'', have received precise answers.

The OWASP list puts these risks at the same impact level.
However, our analysis shows that the impact in the first example is likely to be higher due to the universally acknowledged sensitivity of location data and the potential worst-case outcomes.
This illustrates the additional insights created by our separate analysis of factors for privacy risk.

\section{Conclusion}
This paper set out a research agenda of assessing privacy risk through decomposing privacy risk into separate
factors for both impact and likelihood. We showed how these can be used on relatively coarse ordinal scales,
and illustrated how this can already be used to achieve better insight into specific privacy risks.

The next step would be to refine these metrics, measuring the factors directly, or through proxy measures -- into finer-grained and potentially rational scales; and to look at ways of integrating such metrics that recombine the various dimensions into single values. Inspiration for this may be found in research on multi-dimensional optimization. Such recombination of dimensional metrics becomes essential for several of the potential uses of privacy risk measurement that we indicated above.

The spectrum of metrics that may arise from such refinements and combinations of elementary measurements is likely to be rich. This means that {\em validation} of the alternatives becomes essential, in the first place through considering multiple extensive scenarios with rich collections of privacy risks, for example in the contexts of smart cities or educational data analytics. It has been shown that the strength of privacy metrics can differ between scenarios, and that many metrics have weaknesses at least in some scenarios \cite{wagner2017evaluating}.

There are also mathematical criteria for evaluating privacy metrics. One of these is monotonicity, i.e. that metrics should indicate lower privacy for stronger adversaries \cite{wagner2017evaluating}.
In addition, it may be helpful to calibrate new privacy risk metrics against a database of cases with known privacy risk, for example past cases where the impact is not speculative anymore, in particular with regard to privacy expectation and non-financial harm.

Finally, as with all rigorous methods supporting systems development, we should also take an economical aspect
into account. In privacy risk measurement, we should avoid the false economy of accuracy, noting that ``the time cost of accuracy quite often outweighs the benefits for the organization'' \cite{calderwatkins2015}.
The \ac{GDPR} should increase the uptake of privacy impact assessment in general, but it should not lead to a perception of the process as so complex that it becomes a compliance tool for which cutting corners is desirable.

\section*{Acknowledgment}
This work was supported by the UK Engineering and Physical Sciences Research Council (EPSRC) grant EP/P006752/1.
We thank Lee Hadlington, Richard Snape, and the expert participants of our workshop on ``Privacy risk: harm, impact, assessment, metrics'' in January 2018 for their thoughts and discussions on the topic of this paper.

\bibliographystyle{splncs03}
\bibliography{references}

\end{document}